\documentclass[11pt]{article}%
\usepackage{amsfonts}
\usepackage{amssymb}
\usepackage{graphicx}
\usepackage{setspace}
\usepackage[round]{natbib}
\usepackage{amsmath}%
\usepackage{amsthm}%
\usepackage{bbm}
\usepackage{palatino}
\usepackage{paralist}
\usepackage{multirow}
\usepackage{booktabs}
\usepackage{dsfont}
\usepackage{indentfirst}
\usepackage{enumerate}
\usepackage{enumitem}
\usepackage{longtable}
\usepackage{bm}
\usepackage{caption}
\usepackage{subcaption}
\usepackage{mwe}
\usepackage{float}
\usepackage{booktabs}
\usepackage{lscape}
\usepackage[hyperindex,breaklinks]{hyperref}
\usepackage{hyperref}
\usepackage[utf8]{inputenc}
\usepackage[english]{babel}
\usepackage{skak}
\usepackage{amsmath}

\hypersetup{colorlinks=true,       
	linkcolor=red,       
	citecolor=blue,        
	filecolor=magenta,      
	urlcolor=cyan           
}  

\usepackage{xcolor,csquotes,bm}
\usepackage{tikz}

\usepackage{tikz}
\usetikzlibrary{shapes.misc}

\theoremstyle{definition}

\title{Gambits: Theory and Evidence}

\author{		
         \makebox[.2\linewidth]{Shiva Maharaj\footnote{Email: problemsolver2020@gmail.com}}\\
	\textit{\small  ChessEd}
	\and
	\makebox[.2\linewidth]{Nick Polson\footnote{Email: ngp@chicagobooth.edu}}\\
	\textit{\small  Booth School of Business}\\
	\textit{\small  University of Chicago}\\
        \and
        \makebox[.2\linewidth]{Christian Turk\footnote{Email: cturk23@andover.edu}}\\
	\textit{\small  Phillips Academy}\\
	\textit{\small Andover, MA}\\
	}

\graphicspath{{../fig/}{../wine/}}

\begin{document}
\maketitle


\begin{abstract}
\noindent Gambits are central to human decision-making. Our goal is to provide a theory of Gambits.  A Gambit is a combination of psychological and technical factors designed to disrupt predictable play. Chess  provides an environment to study gambits and behavioral game theory.  Our theory is based on the Bellman optimality path for sequential decision-making. This allows us to  calculate the $Q$-values of a Gambit where material (usually a pawn)  is sacrificed for dynamic play. 
On the empirical side, we study the effectiveness of a number of popular chess Gambits.  This is a natural setting as chess Gambits require a sequential  assessment of a set of moves (a.k.a. policy) after the Gambit has been accepted.   Our analysis uses  Stockfish 14.1 to calculate the optimal Bellman $Q$ values, which fundamentally measures if a position is winning or losing. To test whether Bellman's equation holds in play, we estimate the transition probabilities to the next board state via a database of expert human play. This then allows us to  test whether the \emph{Gambiteer} is following the optimal path in his decision-making. Our methodology is applied to the popular Stafford and reverse Stafford (a.k.a. Boden-Kieretsky-Morphy) Gambit and other common ones including the  Smith-Morra, Goring, Danish and Halloween Gambits. We build on research in human decision-making by proving an irrational skewness preference within agents in chess.  We conclude with directions for future research.

\end{abstract}

\newpage

\section{Introduction}

\emph{What is the object of playing a Gambit opening? To acquire a reputation of being a dashing player at the cost of losing a game----Siegbert Tarrasch}

\vspace{0.1in}

Gambits are central to human decision-making. Our goal is to provide a theory of Gambits.  Rational decision-making under uncertainty has a long history dating back to Ramsey (1926) and de Finetti's (1931) original discussion of previsions and the avoidance of sure loss (a.k.a. dutch book).  This principle is central to the theory of maximum (subjective) expected utility which is prevalent in economic theory, see von Neumann and Morgenstern (1944) and Savage (1954) and Edwards, Lindman and Savage (1963).  Behavioral  decision-making seeks to catalogue all the "anomalies" in human decision-making. The paper focuses on Gambits in chess, when a player offers a material advantage of a pawn (sub-optimal play) for dynamic play in the continuation of the game. 
 Our paper has four goals.  First, we utilize Bellman's principle of optimality and $Q$-learning together with the latest chess engine, Stockfish $14.1$,  to determine, in a concrete way, the rationality of Gambits.  Second, we propose a method of ranking Gambits based on a player's preference for skew and risk. Third, we advocate a new environment, chess, as a way to study behavioral economics.   Fourth, directions for future research. 

 Our work is closely related to the theory of conflict and  negotiation (Schelling, 1960, 2006), game theory (von Neumann and Morgenstern, 1944), and Coase's theorem on bargaining.  
 Schelling's theory applies to a wide range of applications such as defender-attacker relationships to surprise attacks.  Even the threat of a Gambit can have a 
 psychological effect on the other player. Schelling views the strategy of conflict as a bargaining problem and provides numerous conflict situations in  which an un-intuitive 
 and often self-limiting move helps you win.  
 
 Our theory of Gambits has Schelling's feature.  The Gambit itself is giving up a material advantage, a self-limiting policy, enabling
speed and initiative in future play. A plan gets executed quicker, thus being worth more positionally than materially (a.k.a. pawn sacrifices). 

From a game-theoretical perspective, it is well known that in variable-sum games the actions of others can affect your optimal decisions.  
Our approach can be viewed as an application of Adversarial Risk Analysis (ARA) (Rios, Ruggeri, Soyer and  Wilson, 2020). Schelling (1984) observes: 
“The art of looking at the problem from the other person's point of view, identifying his opportunities and his interests, an art that has traditionally been practiced by diplomats, lawyers, and chess players, is at the center of strategic analysis.”
The distributional probabilities of the defender's next state of play against our gambit, given via a database of human play, is modeled by ARA.  Our data is gathered from chess master games and therefore is not meant to be representative of the average chess player.  In fact, the prevalence and success of Gambits is likely to decrease as player skill increases due to linearly increasing player awareness and stronger player engine preparation. Likewise, with waning opponent player skill, the ability to defend against aggressive and complex play diminishes and the odds of success for the Gambiteer increases. 

A Gambit is a combination of psychological and technical factors designed to disrupt predictable play. Once the Gambit is employed---and accepted---the side that accepts the Gambit must play with extreme accuracy to avoid the numerous pitfalls that follow but enjoys a certain win with optimal play. 
A useful running technical definition of a Gambit policy is then: 

\vspace{0.1in}

\emph{A Gambit policy has sure loss in at least one state of the world. But optimal play thereafter has advantages elsewhere, if optimal play is continued under the Bellman principle of optimality}.

\vspace{0.1in} 

In terms of $Q$-values, which measure the gains to optimal play underlining the  principle of optimality (see Section 2.1), we have a formal definition: 
 
 \vspace{0.1in} 
 
\textit{A Gambit policy $G$, with action $ a_G(s)$ in state $s$,  is one in which the $Q\big{(}s,a_G(s)\big{)}<0$ value is negative given fully optimal opponent play, but positive in all other continuation paths given sub-optimal opponent play, resulting in Gambiteer advantage thereafter.}

\vspace{0.1in} 
 
What has not been studied is the propensity for humans to take on Gambits.  We offer a number of economic and behavioral explanations and detailed calculations using principles from sequential decision-making.  Our goal is to study Gambits from both a theoretical and empirical perspective. In the case of the \emph{Gambiteer}, they are willing to sacrifice a pawn as defending the Gambit is equally, if not harder, than risking the Gambit. This is nicely summarised by the following quote  of Karpov--the 12th World champion:

\vspace{0.1in} 

\emph{It doesn't require much for misfortune to strike in the King's Gambit, one incautious move, and Black can be on the edge of the abyss. ---Anatoly Karpov}

\vspace{0.1in} 

An accepted Gambit transfers over "initiative" to the Gambiteer.  When a player has "initiative," they are in control of the style and pace of the game, like a football team when in possession of the ball.  Although Gambits are now easily refuted by Stockfish 14, they still perform well in human amateur play in part due to this fact. 

 John von Neumann described  chess as  a two-player zero-sum game with perfect information and proved the minimax theorem in 1928. His famous quote about chess being a form of computation has been realised in modern day AI with the use of deep neural networks to calculate $Q$-values in Bellman's equation.  This has a number of important ramifications for human play---in a given position there is an optimal action. There is no "style" of play---just optimal play where one simply optimises the probability of winning and follows the optimal Bellman path of play.

Chess AI was pioneered by  Turing (1948), von Neumann, Shannon (1950), and Botvinnik (1970) who developed algorithms for solving chess.  
Shannon's approach was one of trial and error and ”learning” the optimal policy.
Turing (and Champernowne) valued the pieces marginally and in addition they had the following positional evaluation functions: piece mobility, piece safety, king mobility, king safety, and castling. 
Modern day methods are based on state dependent objective function evaluation via $Q$ learning (a.k.a reinforcement learning).  Solving Chess is a daunting NP-hard computational problem, with the Shannon number, which measures the number of possible board states, being $ 10^{152} $ (with $ 10^{43} $ legal moves). 
A major advance over pure look-ahead calculation engines are deep neural networks which interpolate the value and policy functions from empirical game playing.
For example, AlphaZero uses self-play to allow quick solution paths to be calculated and "learns" chess in less than four hours without any prior knowledge, see Dean et al (2012) and Silver et al (2017) for further discussion.

Our focus is on one particular aspect of chess: Gambits.  The word originates from the Italian word, "Gambetto", which most directly translates to, "the act of tripping".  For the sake of clarity, recall our running definition of a Gambit is \emph{a policy that has sure loss in at least one state of the world.  But advantages elsewhere, if optimal play is continued under the principle of optimality.}  
Gambits are commonly played from beginner to advanced levels of chess, however, many people still hold questions about their optimality. Why Gambits?
Our research has shown that Gambits are irrational under the context of optimal play from the opponent; it is always true that the Q-value before the Gambit is higher than the Q-value during optimal play in the Gambit.  

So why are Gambits some of the most common and deadly openings?  The answer lies in skewness preference and humans' natural tendency towards error when making complex decisions.
There have been many studies in behavioral economics that rely on observational data on human decision making in the context of games (Camerer, 2003). The study of the willingness to accept skewed bets has been of central study; for example, the propensity to play lotteries, or to invest in skewed bets on stock-markets where one typically only has preference for mean-variance strategies
(Harvey and Siddique, 2000).  Camerer (1989) empirically shows that markets mis-price bettors demands for bets on team that are extreme underdogs---a form of skewness preference in performance.
There has been extensive literature showing a human preference for positively skewed odds.  In simpler settings, the rational approach has been applied to  bluffing  in poker (von Neumann and Morgenstern, 1944, Ferguson and Ferguson, 2003).
The key feature of skewness, much the same as in Gambits, is the desire to have an extreme payout (a fast quick win) whilst assuming the risk of a high probability of loss.

From an economic perspective, we argue the \emph{Gambiteer} is willing to forego a small material advantage for positive \emph{skewness} in the resulting distribution of optimal continuation $Q$-values. In this sense, the \emph{Gambiteer} is irrational----the decision maker should only worry about optimising future weighted (by the transition probabilities of future states given current state and action) $Q$-values and not the skewness of the distribution. 
Put simply, our main argument for the nature of Gambits is that the psychological effect of the Gambit exceeds the cost of a lost pawn. Our analysis with computer engines shows that Gambits are not optimal strategies.
There is an opposite effect as well, in one of Kasparov's interviews, he explained that his loss to the DeepBlue algorithm was due to the machines ability to avoid traps that typically work with human players.

At first glance, chess may not seem like an apparent place to study economics and decision-making.  In chess and most economic theory, agents are presumed to be acting in the most rational, stable, manner.  In concurrence with the work of behavioral economists, we discuss our finding of a so-called "anomaly" in human behavior in chess, see Camerer and Thaler (1995).

What makes chess a quintessential environment for behavioral economic study?  Successful behavioral experiments demand two critical aspects at their core to avoid running the risk of being waved off and ignored.  First, they need to be repeatable in order to have large sample sizes on human behavior.  For example, one common line of thought attempting to explain behavioral anomalies follows: if people are  given the chance to perform multiple times, they will learn from their mistakes (Thaler, 2014).  Second, experiments need incentives; stakes.  Uncompromising economists try to claim that if the stakes are raised, anomalies in behavior will disappear.  To the dismay of behavioral economists everywhere, this claim is difficult and costly to disprove since a high stake experiment would require great amounts of funding.  A fruitful behavioral theory will be able to counter these arguments.  

The rest of our paper is outlined as follows. The next subsection provides connections with previous research. Section 2 provides an in depth analysis of rationality and sequential decision-making via stochastic dynamic programming. In particular, we describe how to calculate $Q$ values for a decision-making process and show how the principle of optimality provides a constraint that we can test. Section 3 applies our methodology to the Stafford-Boden-Kieseritzky-Morphy Gambits. Section 4 to  the 
Smith-Morra, Goring, Danish and Halloween Gambits. Section 5 concludes and provides directions for future research. 

\subsection{Connections with Behavioral Game Theory} 

Our work builds on an extensive literature of behavioral decision-making (Tversky, 1972) and game theory, see Barberis and Thaler (2012) and Gigerenzer  and  Brighton (2009). 
While behavioral games are immensely useful they are stylistic models in simplified settings and may not generalise to more complex, highly strategic or highly incentivized settings (Camerer, 1997).
There is little to no work on Gambits and complex sequential decision problems.  Humans are more likely to try Gambits in complex settings.  The advent of optimal chess engine moves and assessments now allows us to study such phenomena. 

Holdaway and Vol (2021) analyze over one billion online chess games for human behaviour. They find that players not only exhibit state-dependent risk preferences, but
also change their risk-taking strategy depending on their opponent and that this effect differs in experts and novices. 
As Camerer observes, "Born or made? I think good investors are like chess players.  You have to know a lot. Partly it's temperament. To be a great investor you need a weird combination of aggression and patience."

Kahneman and Tversky  (2013) describe loss aversion and behavioral biases while Camerer and Ho (2014) discuss models of such behaviour, including cognitive hierarchies and $k$-leveling thinking.
Our work builds on "Fairness equilibrium" which incorporates manners into economics (Rabin 1993). In many markets, politeness pays off.
Schelling famously views conflict as a bargaining problem and chess Gambits seem to be no different, rather paradoxically, in such situations, a party can strengthen its position by overly worsening its own options.
The capability to retaliate can be more useful than the ability to resist an attack. As was known to the Greek philosopher Xenophon, uncertain retaliation is more credible and more efficient than certain retaliation---"we ought to learn from our very position that there is no safety for us except in victory."

We also build upon the foundation provided by Jacob Aagaard, who wrote a particularly relevant section in his book series about applying Daniel Kahneman's "system 1" and "system 2" thinking to chess.  Kahneman distinguishes between system 1----'fast', automatic, and subconscious thought, and system 2----'slow', calculating, effortful, and conscious thought (Kahneman, 2013).  Aagaard applies these principles to chess, where players encounter different types of moves, and discusses their corresponding systems and how they can lead to potential for mistakes (Aagaard, 2018).  

Our proposal is that chess is an environment ripe to study behavioral economics and game theory.  To be deemed acceptable, the chess environment should satisfy the formerly mentioned two critical criteria---repeatability and stakes.  Although not a historically popular choice, in recent years, with the advent of the internet and online chess, modern databases give researchers and players alike access to over 8.4 million  humanly generated games, offering the chess environment as a solution to the counter-argument of uncompromising economists.  These multi-million-game databases provide immense diversity in idea, and more importantly, the first critical aspect, repeatability.  Digital chess databases finally allow us to study human behavior in high and low stake environments.  Variable stakes are provided by tournaments, which act as a built-in stake modulator---state and national championships are high stake, high incentive situations, while online blitz (short time-control) games tend to be lower stake.

\section{Chess AI and  Optimal Play}

\vspace{0.1in} 

\emph{Chess problems are the hymn tunes of mathematics.---G. H. Hardy}

\vspace{0.1in} 

The dynamic programming method known as $Q$-learning breaks the decision problem into smaller sub-problems. Bellman's principle of optimality describes how to do this:

\vspace{0.1in}

\emph{Bellman Principle of Optimality: An optimal policy has the property that whatever the initial state and initial decision are, the remaining decisions must constitute an optimal policy with regard to the state resulting from the first decision. (Bellman, 1957)}

\vspace{0.1in}

Backwards Induction identifies what action would be most optimal at the last node in the decision tree (a.k.a. checkmate). Using this information, one can then determine what to do at the second-to-last time of decision. This process continues backwards until one has determined the best action for every possible situation (a.k.a solving the Bellman equation).

\vspace{0.1in}

\noindent{\bf Chess NNUEs}. First, one needs an objective function.  In the case of chess it is simply the probability of winning the game. Chess engines optimize the probability of a win via Bellman's equation and  use deep learning to evaluate the value and policy functions. 
The value function $V(s)$ is simply the probability of winning (100\% (a certain win) to 0\% (a certain loss). For a  given state of the board, denoted by $s$, the value function is given by 
$$
V(s) = \mathbb{P} \left ( \textrm{winning} | s \right )  .
$$
The corresponding $Q$-value is probability of winning, given policy or move $a$ in the given state $s$, by following the optimal Bellman path thereafter, we write 
$$
Q(s,a) = \mathbb{P} \left ( \textrm{winning} | s , a \right ) .
$$
NN engines like AlphaZero don't use centi-pawn evaluations of a position but we can simply transform from centi-pawns to probabilities as follows:
The Win probability $\mathbb{P} \left ( \textrm{winning} | s \right ) $ is related to centi-pawn advantage $c(s)$ in state $s$ of the board via the identity 
$$w(s) = \mathbb{P} (\textit{winning} | s) = 1/(1+10^{-c(s)/4}) \text{ and } c(s) = 4 \log_{10} \big{(}w(s)/(1-w(s))\big{)}.$$

This in turn can be related to Elo rating (Glickman, 1995). Stockfish 14 simply takes  the probability of winning as the objective function. Hence at each stage $V(s)$ measures the probability of winning.
This is typically reported as a centi-pawn advantage.

Hence this will  allow us to test the rationality of Gambits by measuring the difference between optimal play and Gambit play using the optimal Bellman $Q$-values
weighted by the transition probabilities $ \mathbb{P}( s^\star | s , a) $, estimated from human databases, where $s^\star$ is the defined as the next state of the system. At the beginning of the game, Stockfish 14 estimates that the centi-pawn advantage is  $ c(0) = 0.2 $ corresponding to $\mathbb{P} \left ( \textrm{white winning}  \right )  = 0.526 $. 
As opposed to human databases with potentially limited data on certain positions, another way of estimating transition probabilities of human play is to use Maia.  Maia is a chess engine is a neural network like Alpha-Zero, trained from actual human play from the 8 billion game database on LiChess. It has 9 levels of difficulty corresponding to different levels of elo and is the most accurate engine in predicting human play at every level.  Hence, it could provide more accurate estimates via simulation of Maia games, see McIlroy-Young et al (2020).  We leave this as a direction for future research.  See Maharaj, Polson and Turk (2021) for a comparison of other engine architectures like Stockfish and Leela-Zero.     

The optimal sequential decision problem is solved by $Q$-learning, which calculates the $Q$-matrix (Korsos and Polson, 2014, Polson and Witte, 2015), denoted by $Q(s,a)$ for state $s$ and action $a$. 
The $Q$-value matrix describes the value of performing action $a$ (chess move( in our current state $s$ (chess board position))) and then acting optimally henceforth.
The current optimal policy and value function are given by 
\begin{align*}
V(s) & = \max_a \; Q ( s , a )  = Q\big{(} s , a^\star (s)  \big{)},  \\
 a^\star (s) & = {\rm arg max}_a \;   Q ( s , a ).
\end{align*} 
The policy function simply finds the optimal  map from states to actions and substituting into the $Q$ values we obtain the value function at a given states.

The Bellman equation for $Q$-values---assuming instantaneous utility $u(s,a)$ and the $Q$ matrix is time inhomogeneous---is the constraint 
$$
Q( s , a) = u(s,a)+  \sum_{ s^\star \in S } \mathbb{P}( s^\star | s ,a ) \max_{ a } Q ( s^\star , a ),
$$
where $s^\star $ is the new state and $ \mathbb{P}( (s^\star | s ,a )$ denotes the transition matrix of states that describes 
the probability of moving to new state $ s^\star $ given current state $s$ and action $a$.
The new state is clearly dependent on the current action in chess and not a random assignment. 
Bellman's optimality principle is therefore simply describing the constraint for optimal play as one in which the current value is a sum over all future paths of the 
probabilistically weighted optimal future values of the next state.

Taking maximum value over the current action $a$ yields 
\begin{align*}
V(s) &  =  \max_a  \left \{ u(s,a) +  \sum_{ s^\star \in S } \mathbb{P}( s^\star | s ,a )   V (s^\star) \right \}  \; \; {\rm where} \; \; 
 V (s^\star)  = {\rm max}_a \;   Q ( s^\star , a ) .
\end{align*}
The following definition is useful 

\vspace{0.08in} 

\noindent {\bf Definition:}  In a given state $s$, 
a \emph{Gambit Action} $a_G(s)$ has the following property, in terms of $Q$-values:

\vspace{0.1in}

\emph{$ Q ( a_G (s) , s^\star ) < 0 $   and $ Q ( a^\star(s) , s^\star ) > 0 $ for all future states $s^\star$}.

\vspace{0.1in}

However, in order to better model human behavior, we use human game data archived in chess databases such as Chessbase 15 to get a better understanding of human decision-making and risk management when playing with and against Gambits.  We call the human likelihood of an action, "player probability," namely $ \mathbb{P}( s^\star | s ,a )  $.

\subsection{Ranking Gambits by Q-Values}

Table 1 shows Stockfish 14's evaluation of common Gambits. They yield negative Q-values---as required in our definition.  We "rank" some of our personal Gambit favorites by their initial Q-values.  In another later section we perform the same ranking of a smaller subset using a more rigorous analysis based on skewness of future $Q$-values.

$$\begin{tabular}{|l||*{5}{c|}}\hline

\textbf{Opening}&\makebox[3em]{\textbf{Q-Value}}

\\\hline\hline

Stafford Gambit&-2.56

\\\hline\hline

Halloween Gambit&-1.93

\\\hline\hline

Boden-Kiersitsky-Morphy Gambit&-0.87

\\\hline\hline

King's Gambit&-0.76

\\\hline\hline

Budapest Gambit&-0.75

\\\hline\hline

Blackmar-Deimer Gambit&-0.56

\\\hline\hline

Goring Gambit&-0.35

\\\hline\hline

Danish Gambit&-0.35

\\\hline\hline

Smith-Morra Gambit&-0.32

\\\hline\hline

Evans Gambit&-0.25

\\\hline\hline

Queen's Gambit$^*$&+0.39

\\\hline

\end{tabular}
$$
$$\text{Table 1:  Gambit Q-values:  $Q ( a_G , s^\star ) < 0$}$$

*The Queen's Gambit is ineptly named.  White appears to be sacrificing the c4 pawn by allowing black to take it, however, the material is easily regained following d3 and soon after \symbishop xc4.  This is why it is the only "Gambit" on our list with a positive initial Q-value.  This means that if one opts to play the Queen's Gambit, assuming optimal play throughout the rest of the game by both players, the Gambit will win.  

The previous ranking of Gambits considers only the computer evaluation.  Every Gambit, barring the misnomer Queen's Gambit, was evaluated as unsound according to the computer.  In some sense, the computer is the oracle---as close to being completely rational where every agent understands and knows the most optimal lines.

Our proposal for ranking gambits is to use their skew and volatility.  
$$
\begin{tabular}{|l||*{5}{c|}}\hline

\textbf{Opening}&\makebox[3em]{\textbf{Skew}}&\makebox[3em]{\textbf{Volatility}}

\\\hline\hline

Smith-Morra Gambit Variation 1&-0.20 & 0.056

\\\hline\hline

Halloween Gambit Variation 2&+0.34 & 0.089

\\\hline\hline

Danish Gambit Variation 1&+0.89 & 0.096

\\\hline\hline

Stafford Gambit Variation 1&+0.92 & 0.038

\\\hline\hline

Smith-Morra Variation 2&+1.32 & 0.104

\\\hline\hline

Halloween Gambit Variation 1&+1.37 & 0.065

\\\hline\hline

Reverse Stafford Gambit&+1.39 & 0.089

\\\hline\hline

Stafford Gambit Variation 2&+1.45 & 0.112

\\\hline\hline

Danish Gambit Variation 2&+1.49 & 0.096

\\\hline

\end{tabular}
$$

$$\text{Table 2:  Gambit Skew and  Gambit Volatility:  $Q ( a_G , s^\star ) < 0$}$$

Notably, the Smith-Morra (variation 1) is unique in that it has negative skew. The Smith-Morra, played by white, is a common refutation/attack method used when black opts for the solid Sicilian defense.  It is also one of the most popular and winningest gambits due to it's negative skew.  Similar to the Queen's Gambit, it is well known among players to be a "good" gambit----maybe partly due to the fact that computer chess engines rank the Sicilian defense poorly.  

\section{Testing Optimality of Gambits}

$$\includegraphics[width=0.5\linewidth]{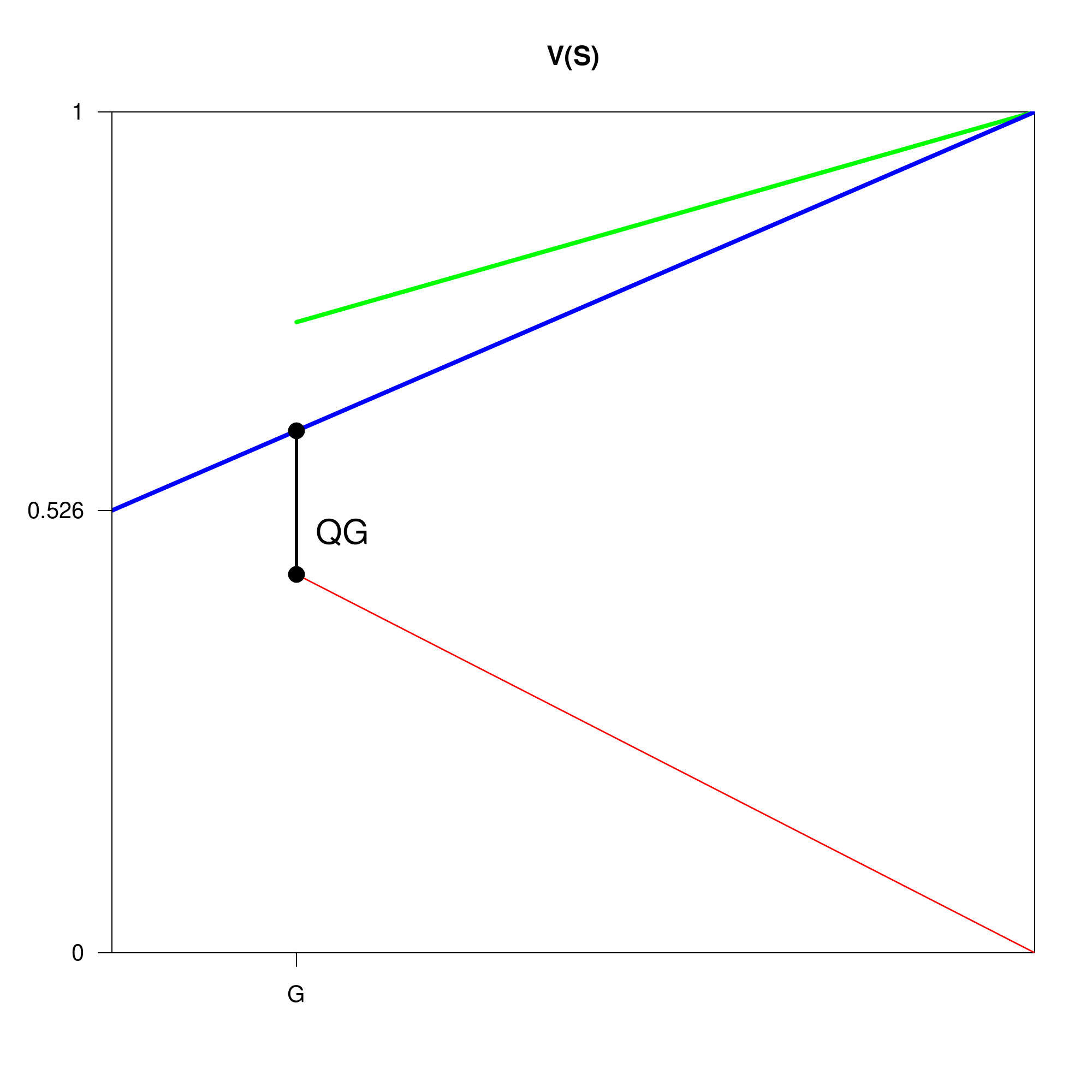}$$
$$\text{Figure 1: Gambit Values $V(G)$ in $[0,1]$ versus time in Game.}$$

Figure 1 illustrates the schematic representation of $Q$-values for a Gambit strategy. The X-axis represents the point in time of the game and the Y-axis represents the objective $Q$-value.  The graph initially starts at $y=0.526$ since chess engines estimate a slightly positive advantage for white, who moves first at the start of the game.  From there, opening moves are played until the point in time, $G$, where the Gambit is played, the Gambiteer offers a material advantage, corresponding to giving up $Q_G$ in $Q$-values.  There are three possible futures starting at point G.  The blue line represents a game where no gambit is offered, or it is declined.  White consistently increases the q-value through extremely optimal play over time and typically wins.  Note that this type of evenly increasing q-value over time is extremely rare, even at super grandmaster levels.  The red line illustrates the future path of the optimal objective function where the opponent, after being challenged to a Gambit, plays optimally, leading to sure loss for the Gambiteer.  The green line illustrates the path of $Q$-values for victory for the Gambiteer, where the opponent accepts the Gambit but does not play optimally due to the complexity and difficulty of playing against a Gambit. 
\bigskip

From our working definition,

\vspace{0.1in}

 \emph{A Gambit policy has sure loss in at least one state of the world. But optimal play thereafter has advantages elsewhere, if optimal play is continued under the principle of Bellman optimality}.

\vspace{0.1in} 

In order to test rationality, we need to calculate 
\begin{enumerate}
\item The value function $V(s)$  (a.k.a. probability of winning) for optimal play. These are simply obtained from Stockfish 14 from a transformed verioin of cent-pawn advantage.
\item  The transition probability matrix $ \mathbb{P}( s^\star | s ,a ) $ where $ s^\star $ is the next state and $s$ the current one. Following the use of ARA we use a  database of human play of games
to estimate these probabilities.
\end{enumerate} 

Then we simply use the  test statistic $T(G)$ defined by the difference between $Q$-values
$$
T (G) = V(s) - Q( s , a_G )  =  Q( s , a_G^* ) - Q( s , a_G ).
$$
This measures how much the \emph{Gambiteer} leaves on the table from playing the Gambit action $a_G $ rather than optimal play $ a^\star ( s ) $. 

The trade-off is the exchange for a distribution of $Q$-values in the next period  given by
$$
\{ Q(  s^\star , a_G ) :  s^\star \in S \; \}. 
$$
Also note that $s$ is the current state of the system, while $s^\star$ is defined as the next state of the system. Looking at the skew of the Q-value distribution is key: by construction, one state has sure loss.    

Restating the Bellman Principle of Optimality as
$$Q(s,a^*(s)) \text{ is optimal} \iff Q\big{(}s^*,a^*(s^*)\big{)}\ge Q\big{(}s,a^*(s)\big{)},$$
allows us to  determine whether playing Gambits is rational by "testing" to see if the equation is satisfied implying  optimal play.

We prove Gambits are sub-optimal since they do not satisfy the Bellman equation.  By opting for a gambit, rather than playing for technical advantage, the Gambiteer expresses a preference for skewness in
the $Q$-values, hinging on error in the next period by the opponent.

Now we apply our understanding of Q-Learning to develop a strategy to analyze and rank Gambits. We will perform our analysis for each of the Gambits based off two "critical moments" where the Gambiteer 
\emph{could} gain an advantage off sub-optimal play from the opponent.

\subsection{Skewness Preferences}

However, in order to more accurately model and understand human Gambit skewness behavior, we will need to adjust our Q-values using transition probabilities from human databases.  First we use the 
aforementioned centi-pawn conversion to arrive at win probability.  Then using human data from a chess database assign probabilities to each move.  
Now we can weigh each actions' win probabilities via their frequency of play, and use our formula for skew to conclude a skewness preference.    

Let  $Q$ denote the random variable that measures the future Q-values at the next time point. 
Skewness is defined via
$$\kappa_G=\sum_{s^*}\mathbb{P}\big{(}s^*|s , a_G(s) \big{)} \left (  \frac{Q(s^*,a_G)-Q^*}{\sigma_G} \right )^3,$$
where  $Q^*$ is  defined as the optimal Q-value from the Bellman equation, namely 
$$Q^*=\sum_{s^*}{\mathbb{P}\big{(}s^*|s, a_G(s)\big{)}V(s^*)} . $$
The volatility of the Gambit strategy is defined by 
$$\sigma_G^2=\sum_{s^*}{\mathbb{P}\big{(}s^*|s, a_G(s) \big{)} \{Q(s^*, a_G)-Q^*\}}^2 . $$
For example, for the Stafford Gambit, using our empirical  data to estimate the state transition probabilities with human play, we find that 
$$\kappa_G(\text{Stafford}) =+0.92.$$ 

For the entire database of gambits, the skew is 
$\kappa_G=+1.70$  with volatility  $\sigma=0.094$.
This aligns with behavioural research on skewness preference for positive skew (Harvey and Siddique, 2000).  

\section{Gambit Evaluations}

\vspace{0.1in} 

\emph{Thinking and planning things will go this or that way---but they are  messy affairs---GM John Nunn}

\vspace{0.1in} 

In the next section we perform analysis of some of the most popular and dubious Gambits.  We use an efficiently updatable neural network (NNUE) chess engine, such as Stockfish 14, to derive current and future Q-values.  Some future Q-values are rooted from Kahneman's "system 1" moves---automatic, fast, and emotional, while others are "system 2" moves---conscious, slow, and analytical.  The former being moves that a player losing concentration might make and the latter being the best moves in the position.

\subsection{Stafford Gambit}

\smallskip

The Stafford Gambit is an opening for black where the Gambiteer allows white to capture a central pawn early on in exchange for extreme piece activity and space.  It has gained popularity among beginner and novice level players due to Twitch.com streamers like Eric Rosen and GothamChess broadcasting themselves playing the Gambit to thousands of live viewers.  

The position below occurs soon after the Gambit is offered by black from the following line of play: 
\newgame
\mainline{1.e4 e5 2.Nf3 Nf6 3. Nxe5 Nc6 4. Nxc6 dxc6 5. d3 Bc5}
$$\showboard$$

$$\text{Figure 2: Stafford Gambit Position 1}$$

$$\begin{tabular}{|l||*{1}{c|}}\hline
Current Q-Value
&\makebox{-2.56}
\\\hline\hline
Pre-Gambit Q-Value
&\makebox{-0.57}
\\\hline\hline
Skew
&\makebox{+0.92}
\\\hline\hline
Volatility ($\sigma$)
&\makebox{0.038}

\\\hline
\end{tabular}$$

$$\text{Table 3:  Stafford Gambit Position 1 Statistics}$$

$$\begin{tabular}{|l||*{5}{c|}}\hline
White \sympawn d3 Move 6    
&\makebox[3em]{\symbishop e2}&\makebox[3em]{\symknight c3}&\makebox[3em]{\symbishop g5}&\makebox[3em]{f3}&\makebox[3em]{\symbishop e3}

\\\hline\hline
Q-value &-2.56&+1.48&+6.20&-1.74&-0.87\\\hline\hline

Player Probability &59\%&4\%&4\%&4\%&14\%

\\\hline\hline
Win Probability &19\%&70\%&97\%&27\%&38\%

\\\hline
\end{tabular}$$

$$\text{Table 4: Stafford Gambit Position 1 Continuations}$$

\smallskip

$$\includegraphics[width=1\linewidth]{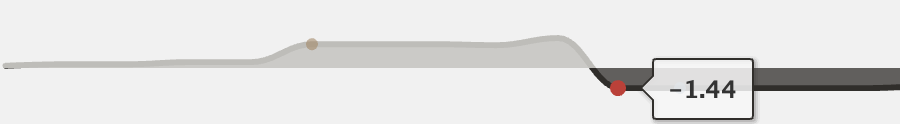}$$
$$\text{Figure 3: Stafford Gambit Q-values over Time}$$

\bigskip

To better understand the relationship between Q-values and time within Gambits, Figure 2 shows the continuation of Q-values for the first 10 moves of the Stafford Gambit.  Much like Figure 1, the graphic starts with a slight positive advantage for white, which steadily increases, until the point of the first yellow dot.  The yellow dot represents when the Gambit is initially offered and, in this case, accepted.  From there, the white player plays optimally until around move 6 where they make a seemingly natural but poor move, resulting in a major blunder and a sharp decrease in Q-value.  At this point, represented by the red dot, the Gambiteer is winning by 1.44 points.

Our methodology displays two tables for each position.  The first contains the Current Q-Value of the position, the Pre-Gambit Q-Value, which is the evaluation prior to the Gambiteer offering the Gambit, the skew ($\kappa_G$) of the data (positive or negative), and finally the volatility ($\sigma$).  The second table displays the raw data of the next five possible states depending on player move.  Player probability represents the likelihood of a real human player to make a move, gathered from data in the database.  Win probability is simply a converted Q-Value to make our data easier to understand.   

Note that for ease of understanding, all of Q-values for this chart have been adjusted to be from the perspective of the Gambiteer.  For example, if the engine reads $-0.50$ and black is the Gambiteer, as is the case within the Stafford Gambit, we adjust the Q-values to show that black is winning, thus turning to a positive $+0.50$.  Looking at the current q-value for example shows that on average, the Gambiteer is approximately 0.93 points down ($-0.93$) in each position we examine.  

\bigskip

An alternative line of play arises from  
\newgame
\mainline{1. e4 e5 2. Nf3 Nf6 3. Nxe5 Nc6 4. Nxc6 dxc6 5. Nc3 Bc5}

$$\showboard$$

$$\text{Figure 4: Stafford Gambit Position 2}$$

$$\begin{tabular}{|l||*{1}{c|}}\hline
Current Q-Value
&\makebox{-2.52}
\\\hline\hline
Pre-Gambit Q-Value
&\makebox{-0.57}
\\\hline\hline
Skew
&\makebox{+1.45}
\\\hline\hline
Volatility ($\sigma$)
&\makebox{0.11}

\\\hline
\end{tabular}$$

$$\text{Table 5: Stafford Gambit Position 2 Statistics}$$

\smallskip

$$\begin{tabular}{|l||*{5}{c|}}\hline
White \symknight c3 Move 6
&\makebox[3em]{h3}&\makebox[3em]{\symbishop c4}&\makebox[3em]{d3}&\makebox[3em]{\symbishop e2}&\makebox[3em]{Qe2}
\\\hline\hline
Q-value &-2.56&+1.43&+1.55&-0.14&-1.62\\\hline\hline
Player Probability &20\%&6\%&7\%&60\%&7\%

\\\hline\hline
Win Probability &19\%&71\%&71\%&48\%&28\%

\\\hline
\end{tabular}$$

$$\text{Table 6: Stafford Gambit Position 2 Continuations}$$

\vspace{0.15in}

The board above displays the initial starting position for the popular Gambit for black against e4, except with an alternative line of play.  For each of the following positions, instead of the current Q-Value in the second half of the principle, we use the "pre-Gambit $Q$-value" as some of the selected moments take place already well into the Gambit, leading to a very negative position.    

In terms of $Q$-values, 
$$-2.55<-0.57,$$
$$Q\big{(}s^*,a^*(s^*)\big{)}<Q\big{(}s,a^*(s)\big{)}.$$
Hence, the Stafford Gambit is highly sub-optimal---although a very popular human choice. 
A simple refutation is \sympawn f3 followed by \sympawn c3.

\subsection{Reverse Stafford Gambit}

The reverse Stafford Gambit (a.k.a. Boden-Kierseritsky-Morphy Gambit) has a win percentage of 62\% one of highest among all white openings.  While it is seldom played by grandmasters because white sacrifices a pawn early on, the inexperienced player can easily make an extremely costly mistake. 

The line of play is given by 
\newgame
\mainline{1. e4 e5 2. Bc4 Nf6 3. Nf3 Nxe4 4. Nc3 Nxc3 5. dxc3}

$$\showboard$$

$$\text{Figure 5: Reverse Stafford Gambit}$$

$$\begin{tabular}{|l||*{1}{c|}}\hline
Current Q-Value
&\makebox{-0.87}
\\\hline\hline
Pre-Gambit Q-Value
&\makebox{+0.62}
\\\hline\hline
Skew
&\makebox{+1.39}
\\\hline\hline
Volatility ($\sigma$)
&\makebox{0.089}

\\\hline
\end{tabular}$$

$$\text{Table 7: Reverse Stafford Gambit Statistics}$$

\smallskip

$$\begin{tabular}{|l||*{5}{c|}}\hline
Black Move 5
&\makebox[3em]{f6}&\makebox[3em]{c6}&\makebox[3em]{d6}&\makebox[3em]{\symbishop c5}&\makebox[3em]{\symknight c6}
\\\hline\hline
Q-value &-0.74&0.00=&+3.63&+2.36&+3.34\\\hline\hline
Player Probability &60\%&14\%&4\%&4\%&4\%

\\\hline\hline
Win Probability &40\%&50\%&89\%&79\%&87\%

\\\hline
\end{tabular}$$

$$\text{Table 8: Reverse Stafford Gambit Continuations}$$

\subsection{Smith-Morra Gambit}

The Smith-Morra Gambit is an opening by white to counter the popular black opening, the Sicillian Defense.  It utilizes a common theme among Gambits, exchanging a pawn for piece activity and space.  
The line of play is given by 
\newgame
\mainline{1.e4 c5 2. d4 cxd4 3. c3 }
$$\showboard$$

$$\text{Figure 6: Smith-Morra Gambit Position 1}$$

$$\begin{tabular}{|l||*{1}{c|}}\hline
Current Q-Value
&\makebox{-0.32}
\\\hline\hline
Pre-Gambit Q-Value
&\makebox{+0.34}
\\\hline\hline
Skew
&\makebox{-0.20}
\\\hline\hline
Volatility ($\sigma$)
&\makebox{0.06}
\\\hline
\end{tabular}$$

$$\text{Table 9: Smith-Morra Gambit Position 1 Statistics}$$

\smallskip

It is worth noting that the Smith-Morra Gambit is the only Gambit we evaluated to have a negative skew value. This is likely because despite White giving up a pawn, the computer evaluation is not significantly against them, meaning in the eyes of the oracle, the exchange for activity in this position is worth it.  Therefore, when players are choosing an opening, the Smith-Morra Gambit (accepted variation shown above) pays off for white assuming slight sub-optimal play from the opponent.  

\smallskip

$$\begin{tabular}{|l||*{5}{c|}}\hline
Black Move 3
&\makebox[3em]{dxc3}&\makebox[3em]{d3}&\makebox[3em]{g6}&\makebox[3em]{\symknight f6}&\makebox[3em]{d5}
\\\hline\hline
Q-value &-0.17&+0.93&+0.90&0.00=&+0.36\\\hline\hline
Player Probability &32\%&11\%&24\%&18\%&4\%

\\\hline\hline
Win Probability &48\%&63\%&63\%&50\%&55\%

\\\hline
\end{tabular}$$

$$\text{Table 10: Smith-Morra Gambit Position 1 Continuations}$$

\bigskip

The line of play for the position below is given by 
\newgame
\mainline{1. e4 c5 2. d4 cxd4 3. c3 dxc3 4. Nxc3 Nc6 5. Nf3 d6 6. Bc4 e6 7. O-O Nf6 8. Qe2 Be7 9. Rd1 Bd7 10. Nb5 Qb8 11.Bf4} 

$$\showboard$$

$$\text{Figure 7: Smith-Morra Gambit Position 2}$$

$$\begin{tabular}{|l||*{1}{c|}}\hline
Current Q-Value
&\makebox{+0.00}
\\\hline\hline
Pre-Gambit Q-Value
&\makebox{+0.34}
\\\hline\hline
Skew
&\makebox{+1.32}
\\\hline\hline
Volatility ($\sigma$)
&\makebox{0.103}

\\\hline
\end{tabular}$$

$$\text{Table 11: Smith-Morra Gambit Position 2 Statistics}$$

\bigskip

$$\begin{tabular}{|l||*{5}{c|}}\hline
Black Move 11
&\makebox[3em]{\symknight e5}&\makebox[3em]{e5}&\makebox[3em]{O-O}&\makebox[3em]{Kf8}&\makebox[3em]{Qd8}
\\\hline\hline
Q-value &0.00=&+1.00&+1.70&+4.64&+5.37\\\hline\hline
Player Probability &12\%&50\%&12\%&12\%&12\%

\\\hline\hline
Win Probability &50\%&64\%&73\%&94\%&96\%

\\\hline

\end{tabular}$$

$$\text{Table 12: Smith-Morra Gambit Position 2 Continuations}$$

\smallskip

Variation 2 of the Smith-Morra Gambit tells a similar story----if black plays optimally, he earns his right to an equal position.  But with one error in concentration, an appearance of Kahneman's 'system 1', the entire position can go south extremely fast for him, resulting in a major advantage to the Gambiteer.  

\bigskip

\subsection{Halloween Gambit}

\smallskip

The Halloween Gambit, also know as the Leipzig Gambit, is an extremely aggressive opening for white where he sacrifices a knight for one of black's central pawns.  We gather two of the most critical moments where a seemingly natural move from black leads to catastrophic failure.  

The line of play is given by 
\newgame
\mainline{1. e4 e5 2. Nf3 Nc6 3. Nc3 Nf6 4. Nxe5 Nxe5 5. d4 Ng6 6. e5 Ng8 7. h4 Bb4 8. h5 N6e7 9. Qg4 g6 10. hxg6 Nxg6 11. Qg3 N8e7 12. Bg5}

$$\showboard$$

$$\text{Figure 8: Halloween Gambit Position 1}$$

$$\begin{tabular}{|l||*{1}{c|}}\hline
Current Q-Value
&\makebox{-0.86}
\\\hline\hline
Pre-Gambit Q-Value
&\makebox{+0.17}
\\\hline\hline
Skew
&\makebox{+1.38}
\\\hline\hline
Volatility ($\sigma$)
&\makebox{0.06}

\\\hline
\end{tabular}$$

$$\text{Table 13: Halloween Gambit Position 1 Statistics}$$

\smallskip

$$\begin{tabular}{|l||*{5}{c|}}\hline
Black Move 12
&\makebox[3em]{O-O}&\makebox[3em]{d6}&\makebox[3em]{d5}&\makebox[3em]{\symknight f5}&\makebox[3em]{\symbishop xc3+}
\\\hline\hline
Q-value &Mate in 5&-0.41&-1.45&-2.43&-0.81\\\hline\hline
Player Probability &20\%&14\%&16\%&38\%&12\%

\\\hline\hline
Win Probability &100\%&44\%&30\%&20\%&39\%

\\\hline

\end{tabular}$$

$$\text{Table 14: Halloween Gambit Position 1 Continuations}$$

\bigskip

The line of play is given by 
\newgame
\mainline{1. e4 e5 2. Nf3 Nc6 3. Nc3 Nf6 4. Nxe5 Nxe5 5. d4 Ng6 6. e5 Ng8 7. Bc4 Bb4 8. Qf3 f6 9. O-O d5 10. exd6 Bxd6 11. Ne4}

$$\showboard$$

$$\text{Figure 9: Halloween Gambit Position 2}$$

$$\begin{tabular}{|l||*{1}{c|}}\hline
Current Q-Value
&\makebox{-0.90}
\\\hline\hline
Pre-Gambit Q-Value
&\makebox{+0.17}
\\\hline\hline
Skew
&\makebox{+0.35}
\\\hline\hline
Volatility ($\sigma$)
&\makebox{0.09}

\\\hline
\end{tabular}$$

$$\text{Table 15: Halloween Gambit Position 2 Statistics}$$

The continuation values are given by 

\smallskip

$$\begin{tabular}{|l||*{5}{c|}}\hline
Black Move 11
&\makebox[3em]{\symknight 8e7}&\makebox[3em]{\symbishop d7}&\makebox[3em]{\symbishop e7}&\makebox[3em]{\symqueen e7}&\makebox[3em]{\symking f8}
\\\hline\hline
Q-value &+1.55&+1.12&-1.17&+1.10&-0.70\\\hline\hline

Player Probability &25\%&6\%&25\%&35\%&6\%

\\\hline\hline
Win Probability &71\%&66\%&34\%&65\%&40\%

\\\hline

\end{tabular}$$

$$\text{Table 16: Halloween Gambit Position 2 Continuations}$$

\bigskip

We now turn to the Danish and Goring Gambits. 

\subsection{Danish and Goring Gambit}

The Goring Gambit arises from White Gambiting the \sympawn d4 after the initial move \sympawn e4.   The Danish Gambit adds one more level to the Gambit by also offering
the \sympawn c3.  If both Gambits are accepted then the Gambiteer has rapid development of his two bishops for the price of losing two pawns. Black neglects his Kingside development and King safety which White then tries to aggressively exploit. 

The line of play is given by 
\newgame
\mainline{1. e4 e5 2. d4 exd4 3. c3 dxc3 4. Nxc3}

$$\showboard$$

$$\text{Figure 10: Danish Gambit Position 1}$$

$$\begin{tabular}{|l||*{1}{c|}}\hline
Current Q-Value
&\makebox{-0.35}
\\\hline\hline
Pre-Gambit Q-Value
&\makebox{+0.40}
\\\hline\hline
Skew
&\makebox{+0.89}
\\\hline\hline
Volatility ($\sigma$)
&\makebox{0.10}

\\\hline
\end{tabular}$$

$$\text{Table 17: Danish Gambit Position 1 Statistics}$$

\smallskip

$$\begin{tabular}{|l||*{5}{c|}}\hline
Black Move 4
&\makebox[3em]{\symknight c6}&\makebox[3em]{\symknight f6}&\makebox[3em]{d6}&\makebox[3em]{\symbishop c5}&\makebox[3em]{\symbishop b4}
\\\hline\hline
Q-value &-0.17&+1.35&0.00=&-0.40&-0.20\\\hline\hline
Player Probability &53\%&3\%&20\%&3\%&20\%

\\\hline\hline
Win Probability &48\%&69\%&50\%&44\%&47\%

\\\hline

\end{tabular}$$

$$\text{Table 18: Danish Gambit Position 1 Continuations}$$

\bigskip

Now consider the following  line of play 
\newgame
\mainline{1. e4 e5 2. d4 exd4 3. c3 dxc3 4. Nxc3 Nc6 5. Bc4 Nf6 6. Nf3 d6 7. Qb3}

$$\showboard$$

$$\text{Figure 11: Danish Gambit Position 2}$$

$$\begin{tabular}{|l||*{1}{c|}}\hline
Current Q-Value
&\makebox{=0.00}
\\\hline\hline
Pre-Gambit Q-Value
&\makebox{+0.40}
\\\hline\hline
Skew
&\makebox{+1.49}
\\\hline\hline
Volatility ($\sigma$)
&\makebox{0.19}

\\\hline
\end{tabular}$$

$$\text{Table 19: Danish Gambit Position 2 Statistics}$$

The table of continuation values is given by 

\smallskip

$$\begin{tabular}{|l||*{5}{c|}}\hline
Black Move 7
&\makebox[3em]{\symqueen d7}&\makebox[3em]{\symbishop e6}&\makebox[3em]{\symbishop e7}&\makebox[3em]{\symqueen e7}&\makebox[3em]{d5}
\\\hline\hline
Q-value &-0.22&+1.97&+2.54&+1.00&+3.36\\\hline\hline
Player Probability &92\%&1\%&1\%&4\%&1\%

\\\hline\hline
Win Probability &47\%&76\%&81\%&64\%&87\%

\\\hline
\end{tabular}$$

$$\text{Table 20: Danish Gambit Position 2 Continuations}$$

\bigskip

Again one can see the skewness in the continuation values.

The final table below shows the overall averages of the gambits and their continuations.

$$\textbf{Summary Table}$$
$$\begin{tabular}{|l||*{1}{c|}}\hline
Current Q-Value
&\makebox{-0.93}
\\\hline\hline
Pre-Gambit Q-Value
&\makebox{+0.14}
\\\hline\hline
Continuation Q-Value
&\makebox{+0.73}
\\\hline\hline
Skew ($\kappa_G$)
&\makebox{+0.99}
\\\hline\hline
Volatility ($\sigma$)
&\makebox{+0.09}
\\\hline\hline
Player Probability
&\makebox{18.9\%}
\\\hline\hline
Win Probability
&\makebox{58.1\%}
\\\hline\hline
Weighted Win Probability
&\makebox{9.38\%}
\\\hline
\end{tabular}$$

$$\text{Table 21: Overall Average Gambit Statistics}$$

The summary table supports the findings of irrational positive skewness preference with an average Gambit skew of +0.99, but only a weighted win probability of 9.38\%.  Weighted win probability was calculated by multiplying the win probability of a given continuation by the player database liklihood of playing that continuation.  This is a more true probability of success for a Gambiteer.  It's also worth noting that on average Gambiteers are winning before the gambit is played by about 0.14 points, but after initiating the gambit, losing by on average -0.93 points.  This furthers the irrational preference for gambits and skew argument.

\section{Discussion}\label{sec:discussion}

We provide a theory of Gambits together with empirical evidence from chess. Surprisingly little academic work has been directed towards Gambits---even though they are an archetypal example of human decision-making.  From a game theory perspective, variable-sum games such as chess are ultra competitive.
For a number of reasons, chess provides a natural empirical setting to study Gambits. We provide a detailed analysis of a number of commonly played Gambits. 
On the one hand, a Gambit policy is not rational; the Gambit leaves open one state of the world where the opponent can win with certainty.  On the other hand, the Gambit leads the Gambiteer to advantage with any sub-optimal play from the opponent.  We provide a number of economic motivations for playing Gambits---from signaling a player type, to a commitment strategy for one to play optimally and dynamically to the theory of conflict and contract theory where surrendering an upside has been shown to be a good negotiation tool.  

From a decision-making behavioral game theory viewpoint, the Gambiteer is expressing a preference for skewness on the Q-Values associated with their optimal path of play.  
Rational planning requires computation of the Bellman optimality path and even the best laid plans can fall by the wayside. "No battle plan ever survives first contact with the enemy"---Helmuth Von Moltke), and in chess strategy, one has to be dynamic.
If the chess market were efficient, Gambits would not be played.  Our view is that while agents have an irrational preference for skew in the Q-values, due to their complexity, Gambits can thrive at all levels of the game, even the most competitive high stakes matches.

From an economic perspective, Tarrasch's view was that the goal of the Gambiteer is not just winning the game but one of also signaling his type to the market of players. His reputation counts.  However, with the increasing prevalence and strength of super computer chess engines, Modern day grandmasters (such as Magnus Carlsen) are more frequently declining Gambits (with no need to let the other player "prove" his policy) thus making themselves unavailable for such threats. We have entered a world of cat and mouse, from the romantic era---of being gentlemanly and accepting Gambits with the burden of proof on the other player. 

There are many directions for future research, both theoretically and empirically. For example, will 
AI provide  new strategies for human chess play and Gambits?  Are there other fields where it will suggest new organizational and decision-making procedures?
Will AI lead to new methods of cooperation, or new methods for the allocation of resources?  In many ways, one human view of AI is that it has by-passed a lot of work that is now unnecessary for humans to perform and opened the door to more leisure time and learning for humans (Keynes, 1930).  
For each gambit, we discuss a player probability derived from the likelihood of a player to make that move based off the database.  It is worth mentioning, however, that there exists a database with 8 billion chess games contained of play for all levels.  Our database was limited to expert play, and thus was only 2 million.  With access to larger databases, or human-like engines like Maia, can we study where exactly humans, not necessarily just experts, are making their mistakes?  Could these mistakes be repeatable and predictable? If so, how can we prevent them? Finally, how can we apply this learning to other areas of decision-making study such as finance?

\section{References}

\noindent Aagaard, J. (2018). \emph{"Grandmaster Preparation: Thinking Inside the Box,"} Quality Chess. \medskip



\noindent Barberis, N. and R. Thaler (2012). "A Survey of Behavioural Finance,"  \emph{Advances in Behavioral Finance Volume}, \emph{III}, pg. 1-76. \medskip 

\noindent Botvinnik, M. (1970).  \emph{Computers, Chess and Long-Range Planning}. Springer-Verlag, NY.\medskip 

\noindent Camerer, C. (1989). “Does the Basketball Market Believe in the ‘Hot Hand’?,” \emph{The American Economic Review 79}, \emph{5}, pg. 1257–1261. \medskip

\noindent Camerer, C. (2003). \emph{Behavioral Game Theory: Experiments in Strategic Interaction}. Russell Sage Foundation.\medskip

\noindent Camerer, C., Loewenstein G., and M. Rabin (2004). "Advances in Behavioral Economics," \emph{Princeton University Press}.   \medskip

\noindent Camerer, C. and T. Ho (2015). "Behavioral Game Theory Experiments and Modeling," \emph{Handbook of Game Theory with Economic Applications}, \emph{4}, pg. 517-573.  \medskip

\noindent Camerer, C. (1997). "Progress in Behavioral Game Theory,"  \emph{Journal of Economic Perspectives,} 11(4), pg. 167-188 \medskip

\noindent Camerer, C. and R. Thaler (1995). "Anomalies: Ultimatums, Dictators, and Manners," \emph{Journal of Economic Perspectives,} 9(6), pg. 209-219.  \medskip

\noindent Dean, J., Corrado, G., Monga, R., Chen, K., Devin, M., Le, Q., Mao, M., Ranzato, M., Senior, A., Tucker, P., Yang, K.,and A. Ng (2012). "Large Scale Distributed Deep Networks," \emph{Advances in Neural Information Processing Systems}, \emph{25}, pg. 1223-1231. \medskip

\noindent de Finetti, B. (1936). "Les Probabilites Nulles," \emph{Bulletin de Sciences Mathematiques}, pg. 275-288.  \medskip 

\noindent Edwards, W., Lindman, H., and J. Savage (1963). "Bayesian Statistical Inference for Psychological Research," \emph{Psychological Review}, \emph{70(3)}, pg. 193-242.  \medskip 

\noindent Ferguson, C. and T. Ferguson (2003). "On the Borel and von Neumann model of Poker," \emph{Game Theory and Applications,} 9, pg. 17-32.\medskip 

\noindent Franchi, F. (2005). "Chess, Games, and Files", \emph{Essays in Philosophy}, \emph{6(1), Article 6}.\medskip

\noindent Gigerenzer, G. and H. Brighton (2009). "Homo Heuristicus: Why Biased Minds Make Better Inferences", \emph{Topics in Cognitive Science},  1(1), pg. 107-143. \medskip

\noindent Glickman, M. (1995). "A Comprehensive Guide To Chess Ratings," \emph{American Chess Journal}, \emph{3(1)}, pg. 59-102. \medskip

\noindent Harvey, C. and A. Siddique  (2000).  "Conditional Skewness in Asset Pricing Tests," \emph{The Journal of Finance}, \emph{55 (3)}, pg. 1263-1295  \medskip 

\noindent Holdaway, C. and E. Vul (2021). "Risk-taking in adversarial games: What can 1 billion online chess games tell us?", \emph{Proceedings of the Annual Meeting of the Cognitive Science Society}, pg. 43. \medskip


\noindent Kahneman D., and A. Tversky (2013). "Prospect Theory: An Analysis of Decision under Risk," \emph{Handbook of Financial Decision Making}, \emph{Part I}, pg. 99-127.  \medskip

\noindent Kahneman D. (2013). "\emph{Thinking, Fast and Slow}," Farrar Straus Girroux.\medskip


\noindent Keynes, J. (1930).  \emph{Economic Consequences for our Grandchildren}. In Essays on Persuasion. \medskip

\noindent Korsos, L. and N. G. Polson (2014). "Analyzing Risky Choices: Q-Learning for Deal-No-Deal," \emph{Applied Stochastic Models in Business and Industry}, \emph{30(3)}, pg. 258-270.\medskip

\noindent Maharaj, S., Polson, N. G., and A. Turk (2021). "Chess AI: Competing Paradigms for Machine Intelligence". \url{arXiv:2109.11602}. \medskip

\noindent McIlroy-Young, R., Kleinberg, J., Siddhartha, S. and A. Anderson (2020).  "Aligning Superhuman AI with Human Behavior: Chess as a Model System", \emph{ACM SIGKDD International Conference on Knoweldge Discovery and Data Mining 2020}, \emph{26}, pg. 1677-1687 \medskip

\noindent Moxley, J., Ericsson, K., Charness, N. and R. Krampe (2012). "The role of intuition and deliberative thinking in experts' superior tactical decision-making", \emph{Cognition}, \emph{vol. 124}, pg. 72-78. \medskip

\noindent Pievatolo, A., Ruggeri, F., Soyer, R., and S. Wilson (2021). "Decisions in Risk and Reliability: An Explanatory Perspective,"  \emph{Stats 2021}, \emph{4}, pg. 228–250. \medskip

\noindent Polson, N.G. and J. Scott (2018). \emph{"AIQ: How People and Machines are Smarter Together"}, St. Martin's press. \medskip

\noindent Polson, N.G. and M. Sorensen (2012). "A Simulation-Based Approach to Stochastic Dynamic Programming," \emph{Applied Stochastics Models in Business and Industry}, \emph{27(2)}, pg. 151-163. \medskip

\noindent Polson, N.G. and J. Witte (2015). "A Bellman view of Jesse Livermore", \emph{Chance}, \emph{28(1)}, pg. 27-31. \medskip

\noindent Rabin, M. (1993). "Incorporating Fairness into Game Theory and Economics," \emph{American Economic Review}, \emph{LXXXIII}, pg. 1281-1302. \medskip

\noindent Ramsey, F. (1926). "Truth and Probability," \emph{The Foundation of Math and Other Logical Essays}. \medskip

\noindent Rios, D., Ruggeri, F., Soyer, F. and S. Wilson(2020). "Advances in Bayesian Decision Making in Reliability", \emph{European Journal of Operational Research}, \emph{282(1)}, pg. 1-18.  \medskip

\noindent Sadler, M. and N. Regan (2019). "Game Changer: AlphaZero's Groundbreaking Chess strategies and the Promise of AI,"  \emph{New in Chess.} \medskip


\noindent Savage, J. (1954). \emph{"Foundations of Statistics."} Wiley. \medskip 

\noindent Schelling, T. (1960).  \emph{"The Strategy of Conflict"}. Harvard University Press.\medskip 

\noindent Schelling, T. (1984).  \emph{"Choice and Consequence"}. Harvard University Press.\medskip

\noindent Schelling, T. (2006), \emph{"Micromotives and Macrobehavior,"}  Norton.  \medskip 

\noindent Shannon, C. E. (1950). "Programming a Computer to Play Chess," \emph{Philosophical Magazine}, \emph{7(41)}, pg. 314.\medskip 

\noindent Shiller, E. (2011). \emph{Attack with the Boden-Kieseritzky-Morphy Gambit}, Chessworks.\medskip

\noindent Silver, D., Huang, A., Maddison, C., Guez, A., Sifre, L., Driessche, G., Schrittweiser, J., Antonoglou, I., Paneershelvam, P., Lanctot, M., Dieleman, S., Grewe, D., Nham, J., Kalchbrenner, K., Sutskever, I., Lillicrap, T., Leach, M., Kavukcuoglu, K., Graepel, T., and D. Hassabis (2017). "Mastering the game of Go without Human Knowledge," \emph{Nature}, \emph{550}, pg. 354-359.\medskip

\noindent Thaler, R. (2014). "\emph{Misbehaving: The Making of Behavioral Economics}," WW Norton. \medskip

\noindent Turing, A. (1948). "Intelligent Machinery," \emph{The Essential Turing}, pg. 395-432. \medskip 

\noindent Tversky A. (1972). "Elimination of Aspects: A Theory of Choice", \emph{Psychological Review}, \emph{79}, pg. 281-299. \medskip

\noindent von Neumann, J. and O. Morgenstern (1944). "The Theory of Economic Games and Behavior," \emph{Princeton University Press}. \medskip 

\end{document}